\documentclass[12pt,aps,prd,preprint,tightenlines,superscriptaddress,showpacs,nofootinbib]
{revtex4-1}
\usepackage{epsfig}
\usepackage{amssymb,amsmath}
\usepackage{color}
\usepackage[utf8]{inputenc} 

\newcommand{\bea}{\begin{eqnarray}}
\newcommand{\eea}{\end{eqnarray}}

 \makeatletter
\def\alt{\mathrel{\mathpalette\gl@align<}}
\def\agt{\mathrel{\mathpalette\gl@align>}}
\def\gl@align#1#2{\lower.6ex\vbox{\baselineskip\z@skip\lineskip\z@
\ialign{$\m@th#1\hfil##\hfil$\crcr#2\crcr\sim\crcr}}} \makeatother

\begin{document}
%\begin{flushright}
%\end{flushright}
%
\vspace*{1.0cm}

\begin{center}
\baselineskip 20pt 
{\Large\bf 
Higgs-Portal Dark Matter in Brane-World Cosmology
}
\vspace{1cm}

{\large 

Taoli Liu$^{~a,}$\footnote{tliu20@crimson.ua.edu}, 
Nobuchika Okada$^{~a,}$\footnote{okadan@ua.edu}, 
and Digesh Raut$^{~b,}$\footnote{draut@smcm.edu}
}
\vspace{.5cm}

{\baselineskip 20pt \it
$^a$Department of Physics and Astronomy, University of Alabama, Tuscaloosa, AL35487, USA \\
$^{b}$ Physics Department, St. Mary's College of Maryland, St. Mary's City, MD 20686, USA
} 

\vspace{.5cm}

\vspace{1.5cm} {\bf Abstract}
\end{center}

The Higgs-portal scalar dark matter (DM) model is a simple extension of the Standard Model (SM)
  to incorporate a DM particle to the SM, where a $Z_2$-odd real scalar field is introduced as a DM candidate.   
We consider this DM model in the context of 5-dimensional brane-world cosmology, 
  where our 3-dimensional space is realized as a hyper-surface embedded in 4-dimensional space. 
In the setup, all the SM and DM fields reside on the hyper-surface while graviton lives in the bulk. 
We consider two well-known brane-world cosmologies, namely, the Randall-Sundrum (RS) 
  and the Gauss-Bonnet (GB) brane-world cosmologies, in which the standard Big Bang cosmology
  is reproduced at low temperatures below the so-called ``transition temperature" while at high temperatures 
  the expansion law of the universe is significantly modified. 
Such a non-standard expansion law directly impacts the prediction for the relic density of the Higgs-portal DM.  
We investigate the brane-world cosmological effects and identify the allowed model parameter region
  by combining the constraints from the observed DM relic density, and the direct and indirect DM detection experiments. 
It is well-known that only DM masses in the vicinity of half the Higgs boson mass are allowed in the Higgs-portal scalar DM model.
We find that the allowed parameter region becomes more severely constrained and even disappears in the RS cosmology, 
  while the GB cosmological effect significantly enlarges the allowed region. 
Upon discovering Higgs-portal DM, we can determine transition temperature in the GB brane-world cosmology. 

\thispagestyle{empty}

%\bigskip
\newpage

\addtocounter{page}{-1}

%%%%%%%%%%%%%%%%%%%%%%%%%%
%\baselineskip 36pt
% Main body
%%%%%%%%%%%%%%%%%%%%%%%%%%
\baselineskip 18pt
%%%%%%%%%%%%%%%%%%%%%%%%%%

%%%%%%%%%%%%%%%%%%%%%%%%
\section{Introduction} 
%%%%%%%%%%%%%%%%%%%%%%%%
According to various astrophysical and cosmological observations, Dark Matter (DM) constitutes about $27\%$ of the energy density of the Universe. 
However, no viable DM particle candidate exists within the Standard Model (SM) of particle physics. 
An electrically neutral weakly interacting massive particle (WIMP) from physics beyond the Standard Model (SM) continues 
  to be an attractive candidate for DM. 
Arguably, the simplest WIMP DM candidate is the so-called Higgs-portal scalar DM \cite{McDonald:1993ex},
  where an SM gauge-singlet scalar added to the SM plays the role of the DM. 
All DM interactions are determined by only two free parameters, namely, the mixed quartic coupling ($\lambda$) 
  between the DM and the SM Higgs doublet and the DM mass ($m_D$). 
To reproduce the observed DM density, the coupling $\lambda$ is determined as a function of $m_D$. 
A combination of constraints from the direct and indirect DM detection experiments as well as the collider search for invisible Higgs decays 
  has already ruled out a large portion of the parameter space of the model, 
  and the DM mass in the vicinity of half of the Higgs boson mass is only allowed for $\lambda$ in perturbative regime.   
See Ref.~\cite{Arcadi:2019lka} for a review on the current status of the Higgs-portal DM scenario. 
Note, however, that this conclusion is based on the standard Big Bang cosmology with three spatial dimensions.

In this paper, we consider the Higgs-portal DM model in the context of the 5-dimensional brane-world cosmology with 4 spatial dimensions. 
In our setup, graviton resides in the whole 4-dimensional space (``bulk"), while all SM and DM particles are confined 
  on the 3-dimensional hyper-surface (``3-brane") embedded in the 4-dimensional space.  
There are two well-known 5-dimensional brane-world cosmologies, namely, the Randall-Sundrum (RS) brane-world cosmology  
  \cite{Binetruy:1999ut,Binetruy:1999hy, Shiromizu:1999wj, Ida:1999ui}
 (for a review, see \cite{Langlois:2002bb}, references therein) and 
 the Gauss-Bonnet (GB) brane-world cosmology 
 \cite{Kim:1999dq,  Kim:2000pz, Nojiri:2002hz, Lidsey:2002zw, Charmousis:2002rc,  Maeda:2003vq}.
At temperatures higher than the so-called transition temperature ($T_t$), the expansion rate of the universe on the 3-brane
  is different from the one of the standard Big Bang cosmology. 
Because the relic density of a WIMP DM depends on the expansion history of the universe, the brane-world cosmological effects
  can significantly alter the resultant DM density from the one in standard Big Bang cosmology. 
For general discussions about the WIMP DM relic density in RS and GB brane-cosmologies, 
  see Refs.~\cite{Okada:2004nc} and \cite{Okada:2009xe}, respectively. 
In what follows, we will investigate the Higgs-portal DM model in the RS and GB brane-cosmologies and 
  identify the allowed parameter region by combining the constraints from the observed DM relic density,  
  and the direct and indirect DM detection experiments.
We will show significant changes from the result in the standard scenario. 
   
%%%%%%%%%%%%%%%%%%%%%%%%%%%%%%%%%%%%%%%%%%%%%%%%%%%%%%%%
\section{The Higgs-portal scalar dark matter model}
%%%%%%%%%%%%%%%%%%%%%%%%%%%%%%%%%%%%%%%%%%%%%%%%%%%%%%%%
%%%%%%%%% Table %%%%%%%%%%%%
\begin{table}[t]
\begin{center}
  \begin{tabular}{|c|ccc|c|}
\hline  
    & $SU(3)_C$ & $SU(2)_L$ & $U(1)_Y$   & ${\bf Z}_2$ \\ \hline
    $q_L^i$ & \bf{3} & \bf{2} & $1/6$  & $+$ \\
    $u_R^i$ & \bf{3} & \bf{1} & $2/3$  & $+$ \\
    $d_R^i$ & \bf{3} & \bf{1} & $-1/3$ & $+$ \\ \hline
    $l_L^i$ & \bf{1} & \bf{2} & $-1/2$   & $+$ \\
    $e_R^i$ & \bf{1} & \bf{1} & $-1$     & $+$ \\ \hline
    $H$ & \bf{1} & \bf{2} & $1/2$        & $+$ \\
    $S$ & \bf{1} & \bf{1} & $0$       & $-$ \\
\hline    
  \end{tabular}
  \renewcommand{\baselinestretch}{1.1}
  \caption{
Particle content of the Higgs-portal DM model. 
The ${\bf Z}_2$-odd real scalar ($S$) is the DM.
} 
\label{Tab:1}  
  \end{center}
\end{table}
%%%%%%%%%%%%%%%%%%%%%%%%%%%%%%%%%%%%%%%%
The particle content of the Higgs-portal DM is listed in Table 1. 
The SM particle content is extended by introducing a $Z_2$ symmetry and a $Z_2$-odd real SM gauge-singlet scalar ($S$). 
All the SM fields are even under the  $Z_2$ symmetry while the scalar $S$ is odd, so that it is stable and serves as the DM. 
The Lagrangian is given as follows: 
\begin{eqnarray}
{\cal L}={\cal L}_{SM}+\frac{1}{2}(\partial_\mu S)(\partial^\mu S)-\frac{m_0^2}{2}S^2-\frac{\lambda_S}{4}S^4
-\lambda S^2  \left( H^\dagger H \right),
\end{eqnarray}
where ${\cal L}_{SM}$ is the SM Lagrangian. 
The DM particle communicates with the SM sector through the ``Higgs-portal" interaction, $\lambda S^2 (H^\dagger H)$. 
After the spontaneous breaking of the electroweak symmetry, we rewrite the above Lagrangian by 
  $H=\frac{1}{\sqrt{2}}\big(\begin{smallmatrix} 0 \\ v_H+h \end{smallmatrix}\big)$ 
  with the Higgs vacuum expectation value $v_H=246$ GeV and the physical Higgs boson ($h$) 
  in the unitary gauge:  
\begin{eqnarray}
{\cal L}={\cal L}_{SM}+\frac{1}{2}(\partial_\mu S)(\partial^\mu S)-\frac{1}{2}m_D^2S^2-\lambda vhS^2-\frac{1}{2}\lambda h^2S^2-\frac{\lambda_S}{4}S^4,
\end{eqnarray}
with  $m_D^2 =m_0^2+\lambda v_H^2$. 
Note that physics of the Higgs-portal DM is controlled by only two free parameters, the DM mass $(m_D)$ and 
 the mixed quartic coupling constant $(\lambda)$. 

%%%%%%%%%%%%%%%%%%%%%%%%%%%%%%%%%%%%%%%%%%%%%%%%%%%
\begin{figure}[t]
\begin{center}
{\includegraphics*[width=0.9\linewidth]{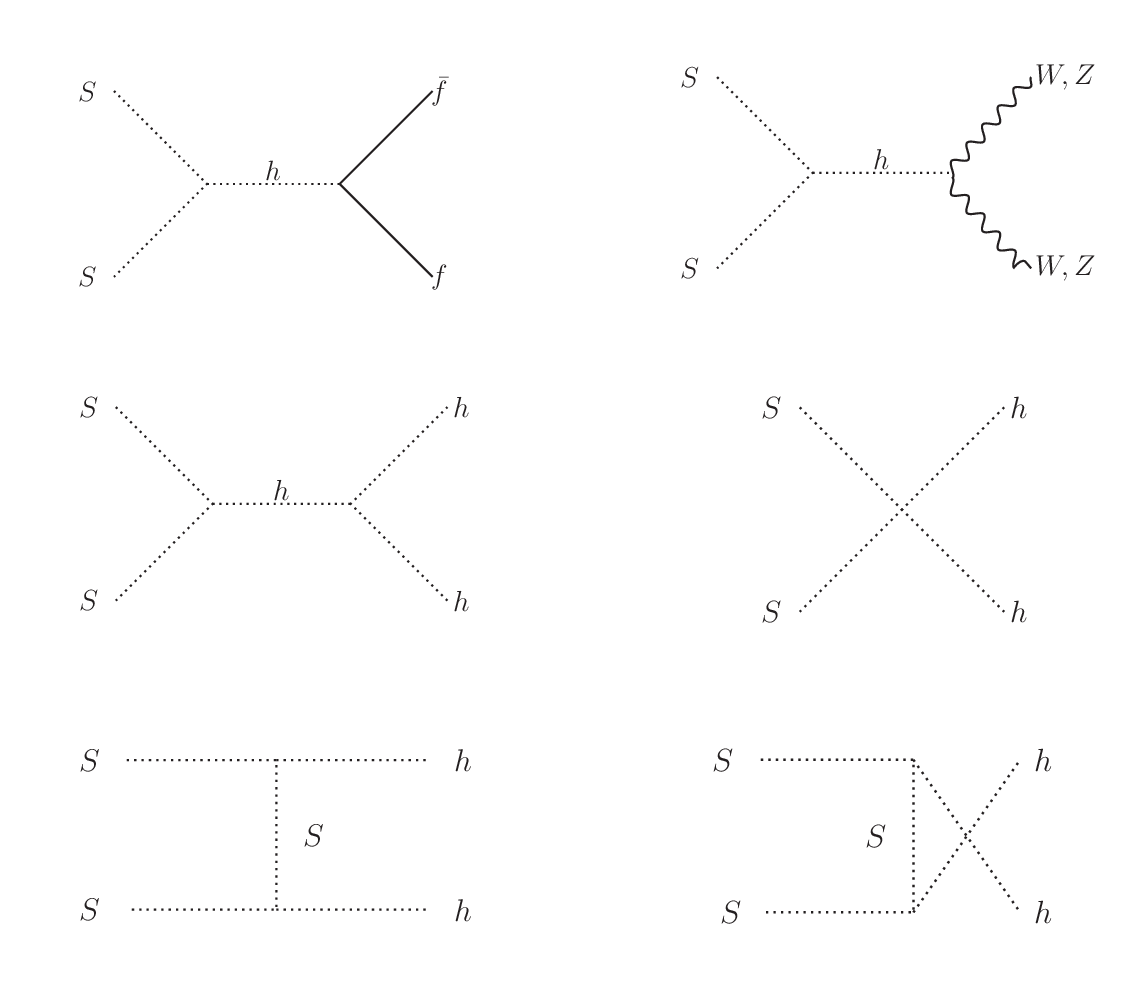}}
\caption{
DM pair annihilation processes to fermion ($f$), anti-fermion ($\bar{f}$), $W$, $Z$ and $h$ bosons. 
}
\label{fig:1}
\end{center}
\end{figure}
%%%%%%%%%%%%%%%%%%%%%%%%%%%%%%%%%%%%%%%%%%%%%%%%%%%

%%%%%%%%%%%%%%%%%%%%%%%%%%%%%%%%%%%%%%%%%%%%%%%%%%%%%%%%%%
\section{Dark matter relic density}
%%%%%%%%%%%%%%%%%%%%%%%%%%%%%%%%%%%%%%%%%%%%%%%%%%%%%%%%%%

To evaluate the thermal relic density of the DM particle, we solve the Boltzmann equation, 
\bea 
  \frac{dY}{dx}
  = - \frac{\langle \sigma v \rangle}{x^2}\frac{s (m_{D})}{H(m_{D})} \left( Y^2-Y_{EQ}^2 \right), 
\label{eq:Boltzman}
\eea  
where $x = m_{D}/T$ and $T$ is the temperature, the DM yield $Y = n/s$ is the ratio of DM number density ($n$) to its entropy density ($s$), 
  $H(m_{D})$ is the Hubble parameter at $T=m_D$, and $Y_{EQ}$ is the DM yield in thermal equilibrium. 
Their explicit forms are 
\bea 
s(m_{D}) = \frac{2  \pi^2}{45} g_\star m_{D}^3,  \; \; 
 H(m_{D}) =  \sqrt{\frac{\pi^2}{90} g_\star} \frac{m_{D}^2}{M_P}, \; \; 
 Y_{EQ}(x) =  \frac{g_{DM}}{2 \pi^2} \frac{x^2 m_{D}^3}{s(m_{D})} K_2(x),   
 \label{eq:st}
\eea 
where $g_\star \simeq 100$ is the total relativistic degrees of freedom of the SM thermal plasma,  
  $M_P=2.44 \times 10^{18}$ GeV is the reduced Planck mass, and 
  $K_2$ is the modified Bessel function of the first kind. 
The thermal average of the DM pair annihilation cross section times relative velocity, $\langle\sigma v\rangle$, in Eq.~(\ref{eq:Boltzman})
  is given by \cite{Edsjo:1997bg}: 
\bea 
\langle \sigma v \rangle =  \frac{g_{DM}^2}{64 \pi^4}
  \left(\frac{m_{D}}{x}\right) \frac{1}{n_{EQ}^{2}}
  \int_{4 m_{D}^2}^\infty  ds \, (\sigma v) \, s \sqrt{s- 4 m_{D}^2} \, K_1 \left(\frac{x \sqrt{s}}{m_{D}}\right),
\label{eq:ThAvgSigma}
\eea
where $g_{DM} = 1$ denotes the degrees of freedom of the real scalar DM $S$,  
$n_{EQ}=s(m_{D}) Y_{EQ}/x^3$ is the thermal equilibrium number density of the DM particle,
$K_1$ is the modified Bessel function of the first kind, and $\sigma (s)$ is the total cross section of DM pair annihilation.  
The DM relic density at the present time is evaluated as
\begin{eqnarray}
\Omega_{DM} h^2=\frac{s_0}{\rho_{crit}^0/h^2}\times m_D \;  Y(x\to\infty),
\end{eqnarray}
where $s_0\simeq2.89\times10^3[{\rm cm}^{-3}]$ is the entropy density at present, 
  and $\rho_{crit}^0\simeq1.05 \, h^2\times10^{-5}[{\rm GeV}/{\rm cm}^3]$ is the critical density
  of the present universe, $h=H_0/(100{\rm km}/{\rm s}/{\rm Mpc})$ is the normalized Hubble constant,
  and $Y_D(x\to\infty)$ is the DM yield at present. 
The DM relic density in the present Universe is $\Omega_{DM} h^2=0.120$ as measured by the Planck 2018 \cite{Aghanim:2018eyx}.

Figure \ref{fig:1} shows the Feynman diagrams for the DM pair annihilation processes to various final SM states, 
  namely, SM fermions ($f$), the weak gauge bosons ($W$ and $Z$) and the SM Higgs boson ($h$). 
The cross sections times relative velocity for these processes are listed below (see, for example, Ref.~\cite{Guo:2010hq}): 
\begin{eqnarray}
{\sigma}_{ff} v &=& \sum_f \frac{\lambda^2 m_f^2}{\pi}
\frac{1}{(s-m_{h}^2)^2+m_{h}^2 \Gamma_{h}^2}
\left( 1-4 \frac{m_f^2}{s} \right)^{\frac{3}{2}}, \label{FF} 
\nonumber \\
{\sigma}_{ZZ} v&=&  \frac{\lambda^2 }{4 \pi}
\frac{s}{(s-m_{h}^2)^2+m_{h}^2 \Gamma_{h}^2} \sqrt{1-\frac{4
m_{Z}^2}{s}} \left(1- \frac{4m_{Z}^2}{s}+ \frac{12 m_{Z}^4}{s^2}
\right),
\nonumber \\
{\sigma}_{WW} v&=& \frac{\lambda^2 }{2 \pi}
\frac{s}{(s-m_{h}^2)^2+m_{h}^2 \Gamma_{h}^2} \sqrt{1-\frac{4
m_{W}^2}{s}} \left(1- \frac{4m_{W}^2}{s}+ \frac{12 m_{W}^4}{s^2}
\right),
\nonumber \\
{\sigma}_{hh} v&=&  \frac{\lambda^2 }{4 \pi s} \sqrt{1-\frac{4
m_{h}^2}{s}} \left[ \left(\frac{s+ 2 m_{h}^2}{s - m_{h}^2}\right)^2
- \frac{16 \lambda v_H^2}{s-2 m_{h}^2} \frac{s+ 2 m_{h}^2}{s
- m_{h}^2} F(\alpha) 
\right. 
\nonumber \\
&& \left. 
+ \frac{32 \lambda^2 v_{H}^4}{(s-2
m_{h}^2)^2} \left( \frac{1}{1-\alpha^2} + F(\alpha)\right) \right],
\label{hh}
\label{eq:DMcs}
\end{eqnarray}
where $m_h = 125$ GeV is the SM Higgs boson mass, $s$ is the square of the center-of-mass energy, 
$F(\alpha)\equiv\mbox{arctanh}(\alpha)/\alpha$ with $\alpha
\equiv\sqrt{(s-4m_{h}^2)(s-4m_D^2)}/(s-2m_{h}^2)$, and 
$\Gamma_{h}$ is the total decay width of the SM Higgs boson, 
including $h\to S S$ if kinematically allowed ($m_h > 2 m_D$),  
\bea
\Gamma_{h} & = & \frac{\sum_f m_f^2 }{8 \pi v_{\rm H}^2}
\frac{(m_{h}^2 - 4 m_f^2)^{3/2}}{m_{h}^2} + \frac{m_h^3}{32 \pi
v_{\rm H}^2} \sqrt{1- \frac{4 m_Z^2}{m_h^2}} \left(1-
\frac{4m_{Z}^2}{m_h^2}+ \frac{12 m_{Z}^4}{m_h^4}\right) \\ \nonumber
& + &  \frac{m_h^3}{16 \pi v_{\rm H}^2} \sqrt{1- \frac{4
m_W^2}{m_h^2}} \left(1- \frac{4m_{W}^2}{m_h^2}+ \frac{12
m_{W}^4}{m_h^4}\right) + \Gamma (h\to S S), 
\label{eq:gammah}
\eea
and 
\bea
\Gamma(h\to S S) =  \frac{\lambda^2 v_{\rm H}^2}{8 \pi}
\frac{\sqrt{m_h^2 - 4 m_D^2}}{m_{h}^2}. 
\eea
The total DM annihilation cross section ($\sigma $) is the sum:   
\bea
 \sigma v = {\sigma}_{ff} + {\sigma}_{ZZ} + {\sigma}_{WW}+ {\sigma}_{hh}. 
\label{eq:tcs}
\eea 
Since all the DM annihilation processes are controlled by only two free parameters, $m_D$ and $\lambda$,  
   we determine $\lambda$ as a function of $m_D$ by imposing $\Omega_{DM} h^2=0.120$.

%%%%%%%%%%%%%%%%%%%%%%%%%%%%%%%%%%%%%%%%%%%%%%%%%%%
\begin{figure}[t]
\begin{center}
{\includegraphics*[width=0.9\linewidth]{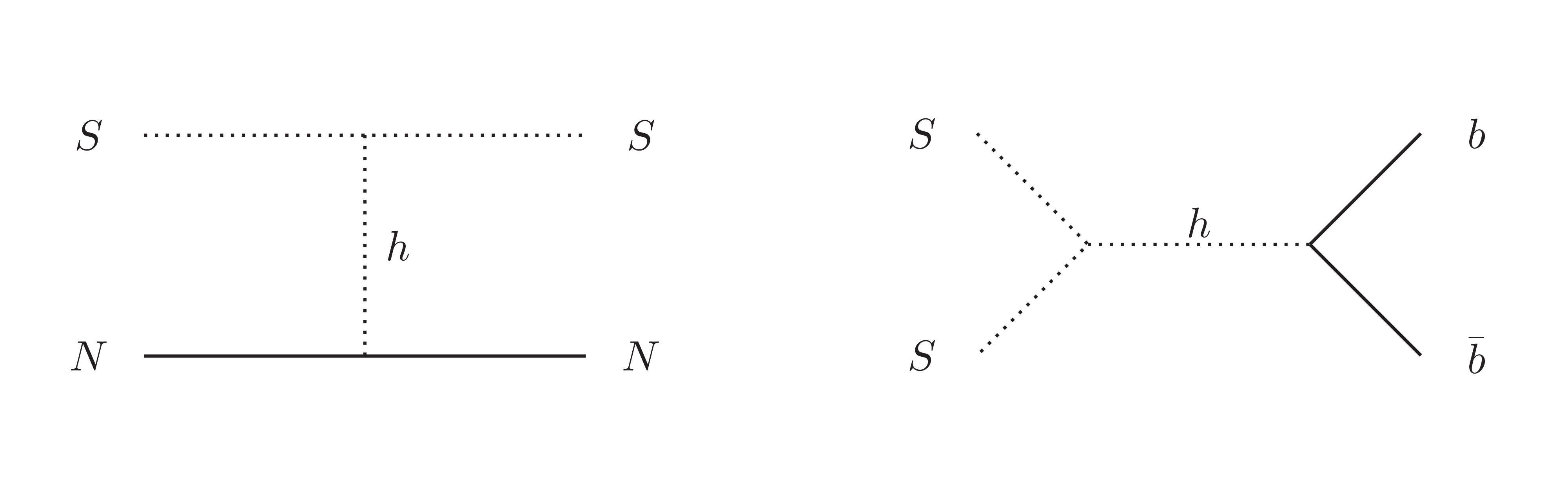}}
\caption{
Left: Feynman diagram for the elastic scattering of DM with nucleon ($N$). 
Right: Feynman diagram for the DM pair annihilation to $b \bar{b}$
}
\label{fig:2}
\end{center}
\end{figure}
%%%%%%%%%%%%%%%%%%%%%%%%%%%%%%%%%%%%%%%%%%%%%%%%%%%

%%%%%%%%%%%%%%%%%%%%%%%%%%%%%%%%%%%%%%%%%%%%%%%%%%%%%%%%
\section{Direct and indirect DM searches}
%%%%%%%%%%%%%%%%%%%%%%%%%%%%%%%%%%%%%%%%%%%%%%%%%%%%%%%%
Various direct DM detection experiments have been searching for elastic scatterings of DM particles off with nucleons. 
The Feynman diagram for this process is shown in the left panel of Fig.~\ref{fig:2}, 
    where $N$ stands for a nucleon (proton or neutron), and $f_N$ is an effective coupling of the DM particle with $N$. 
The spin-independent elastic scattering cross section of the DM particle with a nucleon
   is given by (see, for example, \cite{Kanemura:2010sh})
\begin{eqnarray}
\sigma_{SI}=\frac{\lambda^2}{4m_h^4}
 \left(\frac{m_N}{m_D+m_N}\right)^2 \frac{f_N^2}{\pi}, 
\label{eq:SI}
\end{eqnarray}
where the effective nucleon coupling is calculated as $f_N=0.250$ GeV.
To constrain the model parameters, we adopt the first dark matter search results 
  from the LUX-ZEPLIN (LZ) experiment \cite{LZ:2022lsv}, 
  which provides us with the upper bound on the spin-independent cross section as a function of the DM mass.
As discussed in the previous section, the coupling $\lambda$ is determined as a function of $m_D$ 
  to reproduce the observed DM relic density. 
Thus, under this condition, we can describe $\sigma_{SI}$ as a function of $m_D$ by using Eq.~(\ref{eq:SI}),
  and hence identify the DM mass region to be consistent with the LZ results.

The DM particles in the halo of our galaxy can annihilate in pair and/or decay to create cosmic rays. 
Various indirect DM search experiments have been searching for an excess of cosmic rays 
  originating from such DM pair annihilations and/or late-time decays.
The Higgs-portal DM pair annihilations can create various SM particles, leading to cosmic gamma rays. 
The Fermi Large Area Telescope (Fermi-LAT) experiment has been searching for gamma rays originating from DM pair annihilations. 
The most severe constraints have been obtained by the search results 
  from Milky Way Dwarf Spheroidal Galaxies \cite{Fermi-LAT:2015att},
  which excludes a WIMP DM scenario for a DM mass $\lesssim 100$ GeV if the main DM pair annihilation mode
   is into a bottom quark pair ($b \bar{b})$. 
We adopt the Femi-LAT results to constrain the cross section of the Higgs-portal DM pair annihilation to $b \bar{b}$ 
   since this is the dominant annihilation mode for $m_D \lesssim m_W$. 
For the process shown in the right panel of Fig.~\ref{fig:2}, 
   the thermally averaged cross section at the present time is given by 
%%%%%%%%%%%%%%%%%%%%%%% indirect detect eq. %%%%%%%%%%%%%%%%%%%%%%%%%%%%%%%%%
\begin{eqnarray}
\langle\sigma v\rangle_0= \lim_{x\to\infty}\frac{g_D^2}{(n_D^{eq})^2}\frac{m_D}{64\pi^4x}\int_{4m_D^2}^\infty ds\  
\left( \sigma_{b\bar{b}} v \right) \, s \sqrt{s-4m_D^2} \, K_1\left(\frac{x\sqrt{s}}{m_D}\right),
\end{eqnarray}
%%%%%%%%%%%%%%%%%%%%%%%%%%%%%%%%%%%%%%%%%%%%%%%%%%%%%%%%%%%%%%%%%%
where $\sigma_{b\bar{b}}$ is the DM pair annihilation cross section into a bottom quark pair. 
This cross section is essentially the same as $\sigma_{b\bar{b}} v$ with $s=4 m_D^4$ and $v \to 0$. 
Again, under the constraint of $\Omega_{DM}h^2=0.120$, $\sigma_{b\bar{b}} v$ is uniquely determined
  as a function of $m_D$. 
Therefore, we can identify the allowed DM mass region from the Fermi-LAT results.

%%%%%%%%%%%%%%%%%%%%%%%%%%%%%%%%%%%%%%%%%%%%%%%%%%%%%%%%%%
\section{Brane-world cosmologies}
%\label{sec:2}
%%%%%%%%%%%%%%%%%%%%%%%%%%%%%%%%%%%%%%%%%%%%%%%%%%%%%%%%%%
In this section, we briefly review two typical brane-world cosmologies: the RS and the GB brane-world cosmologies. 
In both scenarios, the standard Big Bang cosmology in 4-dimensional space-time is reproduced 
  at low temperatures while the expansion law of the universe significantly alters at high temperatures. 
We may parametrize a modified Friedmann equation in the brane-world cosmologies
  (non-standard cosmology, in general) as \cite{Okada:2009xe}
\bea
   H(x) = H_{\rm st}(x)  \times F(x), 
\eea
where $H_{\rm st} =H_{\rm st}(m_D) \, x^{-2}$ is the Hubble parameter in the standard Big Bang cosmology (see Eq.~(\ref{eq:st})), 
  and $F(x)$ is a function to represent the modification of the expansion law. 
By introducing the so-called transition temperature ($T_t$) \cite{Okada:2004nc}
  at which the modified expansion law approaches the standard Big Bang expansion law, 
  we may approximately express the function $F(x)$ as 
\bea 
    F(x  <  x_t) &=& \left(\frac{x_t}{x} \right)^\gamma,   \nonumber \\
    F(x > x_t)  &=&1 ,
\label{eq:F}    
\eea
  where $x_t=\frac{m_D}{T_t}$, and $\gamma$ is a real parameter. 
This parameterization turns out to be a very good approximation for both of the RS and GB brane-world cosmologies. 
In the following, we will see that $\gamma= 2$ corresponds to the RS brane-world cosmology 
   while $\gamma= -2/3$ to the GB brane-world cosmology.

%%%%%%%%%%%%%%%%%%%%%%%%   
\subsection{RS brane-world cosmology}
%%%%%%%%%%%%%%%%%%%%%%%%   
The RS brane-world cosmology is based on the action given by \cite{Randall:1999vf}
\bea
 {\cal S} = \frac{1}{2\kappa_5^2} \int d^5x 
 \sqrt{-g_5}\left[ - 2 \Lambda_5+ {\cal R} \right]
- \int_{brane} d^4x 
 \sqrt{-g_4} \left( m_\sigma^4 + {\cal L}_{matter} \right), 
\label{RSaction}
\eea
where $\kappa_5^2=8\pi/M_5^3$ with a 5-dimensional Planck scale $M_5$, $ m_\sigma^4 >0 $ is a brane tension, 
  $ \Lambda_5 <0 $ is a negative bulk cosmological constant, 
  and a $Z_2$-parity across the brane in the bulk is imposed. 
The Friedmann equation for a spatially flat universe on the brane 
  is found to be \cite{Binetruy:1999ut,Binetruy:1999hy, Shiromizu:1999wj, Ida:1999ui} 
\begin{equation}
H^2 = \frac{\rho}{3 M_P^2} \left(1+\frac{\rho}{\rho_{\rm RS}} \right) ,
\label{Eq.RSH}
\end{equation}
where $\rho$ is the energy density on the brane, and
\begin{eqnarray}
 \rho_{\rm RS} = 12 \,  \frac{M_5^6}{M_P^2}. 
\label{rho_0}
\end{eqnarray}
Here, we have set the model parameters so as to vanish the 4-dimensional cosmological constant 
   and the so-called dark radiation \cite{Ichiki:2002eh}.  
Note that the Friedmann equation of the standard Big Bang cosmology is reproduced 
    at low energies (temperatures), namely $\rho/\rho_{\rm RS} \ll 1$.  
%A lower bound on $\rho_{\rm RS}^{1/4} \gtrsim 1.3$ TeV, or equivalently, $M_5 \gtrsim 1.1 \times 10^8$ GeV
%   was obtained in Ref.~\cite{Randall:1999vf} from the precision measurements of the gravitational law in sub-millimeter range. 
In the radiation dominated era with a high temperature $\rho/\rho_{\rm RS} \gg 1$. 
  the Friedmann equation of  Eq.~(\ref{Eq.RSH}) can be approximated by
\bea
  H   \simeq H_{\rm st} \, \sqrt{\frac{\rho}{\rho_{\rm RS}}} = H_{\rm st} \times \left( \frac{x_t}{x} \right)^2
\eea 
   where we have used $\rho/\rho_{\rm RS} = (T/T_t)^4 = (x_t/x)^4$. 
Thus, the RS brane-world cosmology corresponds to $\gamma=2$ in Eq.~(\ref{eq:F}) for $T \gg T_t$.

%%%%%%%%%%%%%%%%%%%%%%%%
\subsection{GB brane-world cosmology}
%%%%%%%%%%%%%%%%%%%%%%%%
The action in Eq.~(\ref{RSaction}) can be generalized by adding higher curvature terms. 
Among various possible terms, the GB invariant is of particular interests in 5-dimensions,  
  since it is a unique nonlinear term in curvature yielding the gravitational field equations at the second order. 
The GB brane-world cosmology is based on the extension of Eq.~(\ref{RSaction}) 
  by adding  the GB invariant  \cite{Kim:1999dq,  Kim:2000pz, Nojiri:2002hz, Lidsey:2002zw}: 
\bea
 {\cal S} &=& \frac{1}{2\kappa_5^2} \int d^5x 
 \sqrt{-g_5}
 \left[
 - 2 \Lambda_5+ {\cal R} + 
 \alpha \left( {\cal R}^2 -4 {\cal R}_{ab} {\cal R}^{ab} 
 + {\cal R}_{a b c d}{\cal R}^{a b c d} \right) \right] \nonumber \\
&-& \int_{brane} d^4x 
 \sqrt{-g_4} \left( m_\sigma^4 + {\cal L}_{matter} \right), 
\label{GBaction}
\eea
where $\alpha$ is a constant, and the indices $a,b,c,d$ run $0$ to $4$. 

The Friedmann equation on the spatially flat brane has been found to be \cite{Charmousis:2002rc,  Maeda:2003vq}
\bea 
 \kappa_5^2(\rho + m_\sigma^4) = 
  2 \mu \sqrt{1+\frac{H^2}{\mu^2}}
  \left( 3 - \beta +2 \beta \frac{H^2}{\mu^2} \right) ,  
\label{GBFriedmann1} 
\eea
where $\beta = 4 \alpha \mu^2 = 1-\sqrt{1 + 4 \alpha \Lambda_5/3}$. 
The model involves four free parameters, $\kappa_5$, $m_\sigma$, $\mu$ and $\beta$. 
The following two phenomenological constraints are imposed: 
(i) requiring that the Friedmann equation of the standard Big Bang cosmology is reproduced at low energies $H^2/\mu^2 \ll 1$, and 
(ii) requiring that 4-dimensional cosmological constant to be zero, 
which lead to the following relations:
\bea 
 \kappa_5^2 m_\sigma^4 = 2 \mu (3-\beta), \; \; \;  \frac{1}{M_P^2} 
  = \frac{\mu}{1+\beta} \kappa_5^2 .
\label{relations} 
\eea
In general, the GB cosmology has three epochs in its evolution \cite{Lidsey:2003sj}: 
The universe obeys the standard expansion law at low energies (standard epoch). 
At middle energies (RS epoch), the RS brane-world cosmology solution is approximately realized. 
At high energies (GB epoch), the Friedmann equation is approximately expressed as
\bea 
 H \simeq 
 \left(  \frac{1+\beta}{4 \beta}  \, \frac{\mu}{M_P^2}  \, \rho  \right)^{1/3}.  
\eea
It is known that the RS epoch collapses for $\beta = 0.151$ \cite{Lidsey:2003sj}, 
  which is the solution to $3 \beta^3 -12 \beta^2 +15 \beta -2 =0$. 
To focus on the GB brane-world cosmology, we consider this case. 
Hence, we can parametrize the Friedman equation in the GB epoch as 
\bea
   H \simeq  \left(  \frac{1+\beta}{4 \beta}  \, \frac{\mu}{M_P^2}  \, \rho  \right)^{1/3} 
    = H_{\rm st}  \times  \left( \frac{\rho}{\rho_{\rm GB}}  \right)^{-1/6}
    = H_{\rm st}  \times  \left( \frac{x_t}{x}  \right)^{-2/3}, 
\eea
   where $\rho_{\rm GB}=3^6 \left(\frac{1+\beta}{4 \beta} \mu M_P \right)^2$. 
Thus, the GB brane-world cosmology corresponds to $\gamma=-2/3$ in Eq.~(\ref{eq:F}) for $T \gg T_t$.

%%%%%%%%%%%%%%%%%%%%%%%%%%%%%%%%%%
\section{Higgs-portal dark matter in Brane-world Cosmologies}
%%%%%%%%%%%%%%%%%%%%%%%%%%%%%%%%%%
Let us now investigate the brane-world cosmological effects on the Higgs-portal DM scenario. 
We first calculate the DM relic density. 
By using the modification of the Hubble parameter in Eq.~(\ref{eq:F}), 
  the Boltzmann equation is given by 
\bea 
  \frac{dY}{dx}
  = - \frac{\langle \sigma v \rangle}{x^2 F(x)}\frac{s (m_{D})}{H_{\rm st}(m_{D})} \left( Y^2-Y_{EQ}^2 \right).  
\label{eq:Boltzman-BWC}
\eea  
Comparing this equation with Eq.~(\ref{eq:Boltzman}), we can see that the brane-world cosmological effect
  is equivalent to modifying the DM annihilation cross section in the standard Big Bang cosmology 
  to an effective one defined as 
\begin{eqnarray}
\langle \sigma v \rangle \to \langle \sigma v \rangle_{\rm eff} = \frac{\langle\sigma v\rangle}{F(x)} \simeq 
 \langle\sigma v\rangle \left( \frac{x}{x_t} \right)^{\gamma} = \langle\sigma v\rangle \left( \frac{T_t}{T} \right)^{\gamma}
 \label{eq:sigmaeff}
\end{eqnarray}
for $T > T_t$. 
As expected, the brane-world cosmological effect is important only when the DM decouples from the SM thermal plasma at $T > T_t$. 
With fixed $m_D $ and $\lambda$ values, the effective cross section is reduced (increased)
  in the RS (GB) brane-world cosmology with $\gamma=2$ ($\gamma=-2/3$) from the one in the standard Big Bang cosmology. 
It is well known that the DM annihilation cross section times relative velocity $\sim 1$ pb is suitable to reproduce 
  the observed DM relic density of $\Omega_{DM} h^2=0.120$. 
This fact indicates that to reproduce the observed DM density for a fixed $m_D$ value, 
  the $\lambda$ value is taken to be larger (smaller) for the RS (GB) cosmology 
  from its value in the standard Big Bang cosmology (see Eq.~(\ref{eq:DMcs})).  

%%%%%%%%%%%%%%%%%%%%%%%%%%%%
\begin{figure}
    \centering
    \includegraphics[width=9cm]{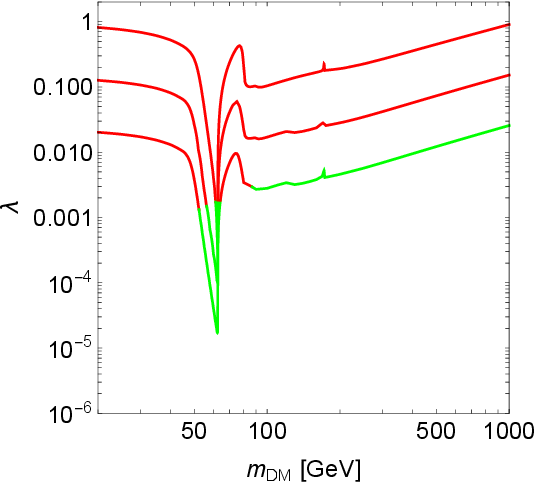}
\caption{
Allowed parameter region in ($m_D, \lambda$)-plane for the standard Big Bang (middle) and brane cosmologies: RS with $x_t = 500$ (top) and GB with $x_t = 10000$ (bottom). 
The observed DM relic density of $\Omega_{DM} h^2=0.12$ is reproduced along the curves. 
The green regions depict the allowed regions for the three cases 
  after imposing the direct and indirect DM detection constraints. 
}
    \label{fig:final}
\end{figure}
%%%%%%%%%%%%%%%%%%%%%%%%%%%%
  
We numerically solve the Boltzmann equation of Eq.~(\ref{eq:Boltzman-BWC}) 
   to evaluate the relic density of Higgs-portal DM in 4-dimensional, RS and GB brane-world cosmologies 
   for three benchmark transition temperature values, $x_t =0, 500,$ and $10000$, respectively. 
Note that the standard Big Bang cosmology corresponds to setting the transition temperature to be infinite 
   or equivalently, $x_t \to 0$. 
We set a model-independent lower bound on the transition temperature to be $T_t > 1$ MeV 
   not to ruin the success of the BBN. 
This corresponds to a bound on $x_t < 10^3 \times m_D[{\rm GeV}]$.
For example, $x_t \lesssim  10^5$ for $m_D \sim 100$ GeV. 
To reproduce the observed DM density of $\Omega_{DM} h =0.120$, we have determined 
   the coupling $\lambda$ as a function of $m_D$. 
We show our results in Fig.~\ref{fig:final}. 
The plots from top to bottom correspond to the RS, the standard Big Bang, and the GB cosmologies, respectively, 
   along which the observed DM density is reproduced. 
As we have expected, the resultant $\lambda$ value for a fixed $m_D$ is larger (smaller) than the one 
  in the standard Bing Bang cosmology for the RS (GB) brane-cosmology. 
The green and red color coding identify the allowed and excluded range of the DM mass after imposing constraints from DM direct and indirect detection experiments, as we will discuss in the following. 

%%%%%%%%%%%%%%%%%%%%%%%%%%%%%%%%%%%%%%%%%%%%%%%%%%%%%%
\begin{figure}[t]
    \centering
    \includegraphics[width = 5 cm]{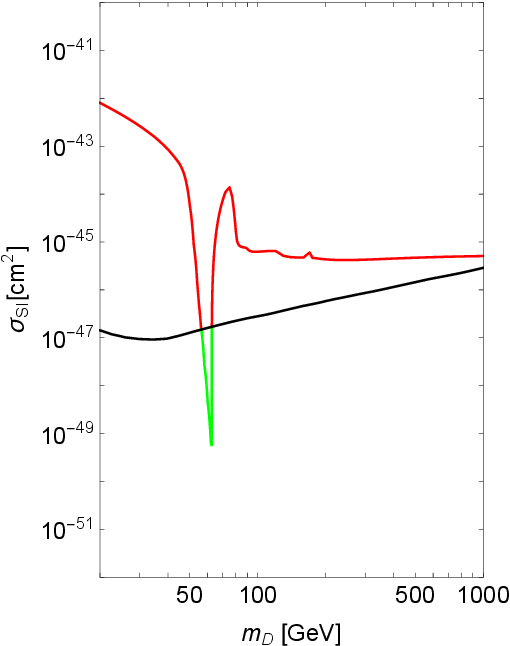}\; \;  \includegraphics[width = 5 cm]{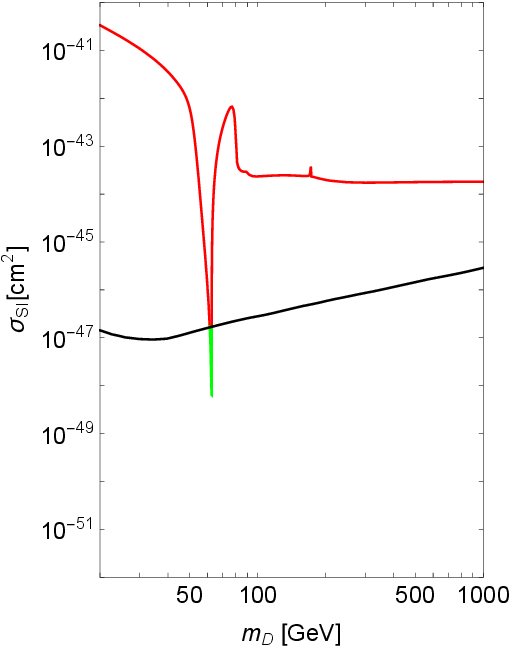}\; \; 
    \includegraphics[width = 5 cm]{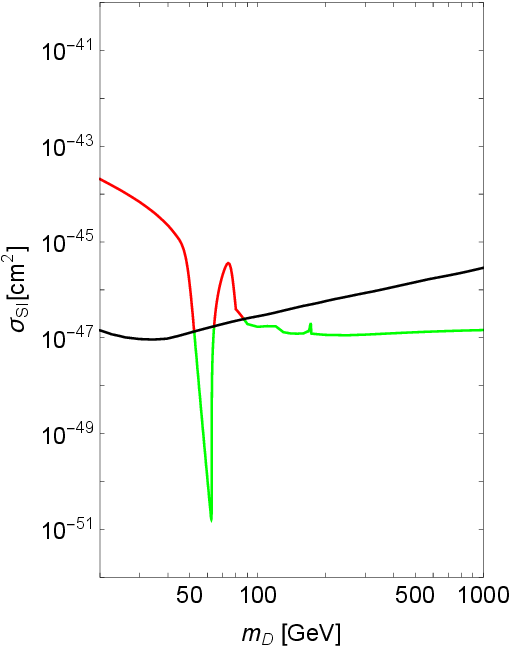}
\caption{
The DM-nucleon elastic scattering cross section as a function of DM mass for the standard Big Bang (left) 
  and brane cosmologies: RS with $x_t = 500$ (middle) and GB with $x_t = 10000$ (right). 
The observed DM relic density of $\Omega_{DM} h^2=0.12$ is reproduced along the curves. 
In each panel, the upper-bound on the cross section from the LZ experiment is depicted by the black solid curve. 
The green regions are allowed. 
}
    \label{fig:direct}
\end{figure}
%%%%%%%%%%%%%%%%%%%%%%%%%%%%

Using the results obtained in Fig.~\ref{fig:final}, we calculate the spin-independent elastic cross section ($\sigma_{SI}$) with a nucleon as a function of $m_D$.   
We show our results in Fig.~\ref{fig:direct} along with the upper bound 
  from the LZ experiment. 
The panels from left to right correspond to the standard Big Bang cosmology, the RS brane-world cosmology with $x_t = 500$, 
   and the GB brane-world cosmology with $x_t = 10000$, respectively. 
The regions in red are excluded since the predicted cross sections exceed the LZ bound (black horizontal lines). 
 As well-known, the allowed parameter region for the Higgs-portal DM model in the standard Big Bang cosmology 
   is very severely constrained. 
We have found that the allowed parameter region becomes more severely constrained in the RS brane-world cosmology. 
As we raise $x_t$ further, the allowed region eventually disappears. 
The result in the GB brane-world cosmology is probably more interesting since this effect enlarges the allowed region, 
  and only the regions of $m_D \lesssim m_h/2$ and $m_D \simeq m_W$ are excluded.

\begin{figure}[t]
    \centering
    \includegraphics[width = 5 cm]{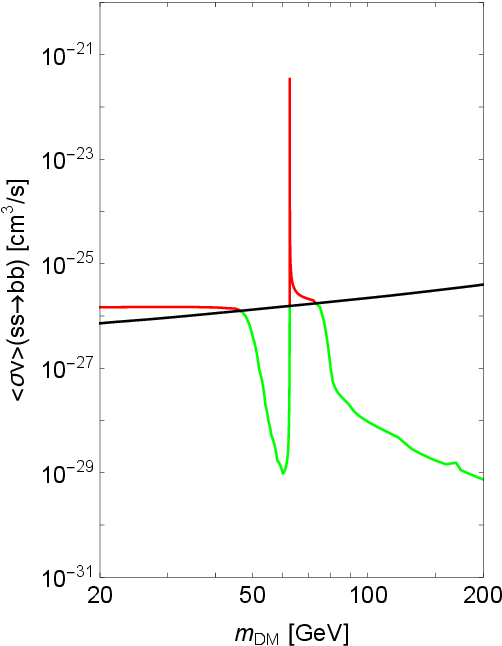}\; \;  \includegraphics[width = 5 cm]{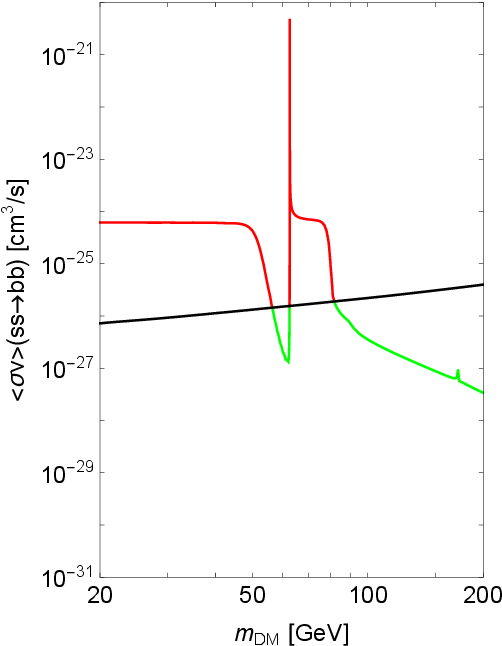}\; \; 
    \includegraphics[width = 5 cm]{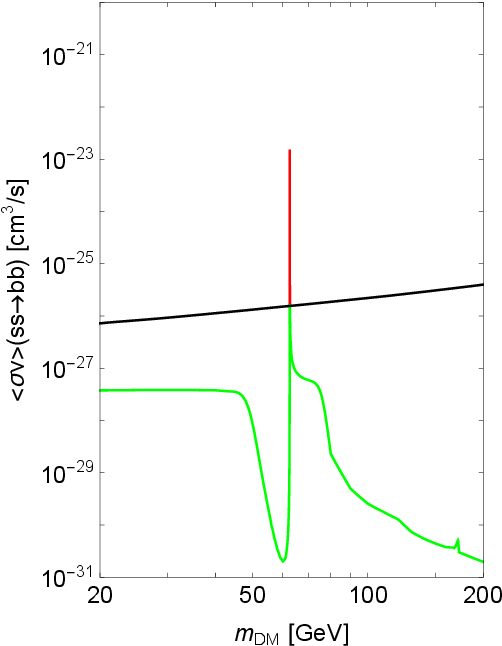}
\caption{
The DM pair annihilation cross section to $b \bar{b}$ as a function of DM mass for the standard Big Bang (left) 
  and brane cosmologies: RS with $x_t = 500$ (middle) and GB with $x_t = 10000$ (right). 
The observed DM relic density of $\Omega_{DM} h^2=0.12$ is reproduced along the curves. 
In each panel, the upper-bound on the cross section from the Fermi-LAT experiment is depicted by the black solid curve. 
The green regions are allowed. 
}
    \label{fig:indirect}
\end{figure}

Using the results obtained in Fig.~\ref{fig:final}, we also calculate the DM pair annihilation cross section to $b \bar{b}$. 
We show our results in Fig.~\ref{fig:indirect} along with the indirect DM detection constraint from the Fermi-LAT result.   
The panels from left to right correspond to the standard Big Bang cosmology, the RS brane-world cosmology with $x_t = 500$, 
   and the GB brane-world cosmology with $x_t = 10000$, respectively.  
The regions in red are excluded since the predicted cross sections exceed the Fermi-LAT bound (black horizontal lines). 
Similarly to the result from the direct DM detection constraint, the parameter region in the RS brane-world cosmology 
   is more severely constrained than the standard Big Bang cosmology case,
   while the indirect detection experiment provides (almost) no constraint on the GB case.

%%%%%%%%%%%%%%%%%%%%%%%%%%%%
%%%%%%%%%%%%%%%%%%%%%%%%%%%%

The results in Fig.~\ref{fig:final} show the allowed parameter region in the ($m_D, \lambda$)-plane after imposing both  direct and indirect DM detection constraints. 
As we have already discussed, the observed DM density is reproduced along the curves for the corresponding cosmologies. 
The green and red color coding identify the allowed and excluded range of the DM mass after imposing the constraints. Three curves from top to bottom correspond to the results for the RS brane-world cosmology with $x_t = 500$, 
  the standard Big Bang cosmology, and the GB brane-world cosmology with $x_t = 10000$, respectively.    
We have found very narrow allowed regions for the RS and the standard Bing Bang cosmologies 
  while in the GB brane-world cosmology, the large portion of the parameter space for $m_D \gtrsim m_h/2$ is still allowed. 
For $m_D < m_h/2$, the Higgs boson can invisibly decay to a pair of the Higgs-portal DMs.  
Thus, the Higgs-portal DM model is also constrained by the (null) results from the search for invisible decays
  of the Higgs boson at the Large Hadron Collider (LHC). 
Since this constraint is much weaker than the one from the LZ experiment for the range of $m_D > 20$ GeV \cite{ATLAS:2023tkt}, 
  we did not show it in this paper. 
However, the LHC results provide us the most severe constraints for $m_D \lesssim 5$ GeV.

%%%%%%%%%%%%%%%%%%%%%%%%%%%%%%%%%
\section{Conclusion}
%%%%%%%%%%%%%%%%%%%%%%%%%%%%%%%%%
Although one of the primary candidates for the Dark Matter (DM) particle in the Universe is the electrically neutral 
   and stable WIMP, such a particle is absent in the SM.  
The Higgs-portal scalar DM model is a simple extension of the SM to supplement the SM with a WIMP DM candidate, 
   a $Z_2$-odd real scalar. 
Higgs-portal DM particle communicates with the SM sector exclusively through its coupling with the SM Higgs doublet. 
Consequently, the physics of the Higgs-portal DM is governed by only two free parameters: 
   the DM mass ($m_D$) and the mixed quartic coupling ($\lambda$). 
To reproduce the observed DM relic density, the coupling $\lambda$ is determined as a function of $m_D$. 
Hence all the predictions of the model can be determined as a function of the DM mass. 
Since this model is very predictive we can scrutinize it through various experiments and observations. 
After combining constraints from the observed DM relic density, direct and indirect DM detection experiments, 
  and the LHC search for invisible Higgs boson decays, it becomes evident that almost the entire parameter space is excluded: only $m_D$ values in the vicinity of half of the Higgs boson mass is allowed.

The brane-world scenario proposes an interesting possibility that our 3-dimensional space is realized
   on a hyper-surface embedded into 4 or more spatial dimensions. 
In this setup, there are two phenomenologically viable cosmologies in five (=1+4) dimensions, 
   the RS and the GB brane-world cosmologies, 
   which can reproduce the standard Big Bang cosmology on our 3-dimensional universe at low energies 
   (temperatures below the transition temperature) 
   while the expansion law of the universe at high energies are significantly altered. 
We have considered the Higgs-portal DM model in the brane-world cosmologies 
   and investigated the brane-world cosmological effects on the model. 
If the DM decouples from the SM thermal plasma when the universe undergoes non-standard expansion, 
   the resultant DM relic density can be significantly altered. 
   Hence the coupling $\lambda(m_D)$ to reproduce 
   the observed DM density differs significantly from the one obtained in the standard Big Bang cosmology. 
The difference becomes wider as the transition temperature decreases. 
Requiring that the WIMP DM to reproduce the observed abundance of DM, we have calculated the spin-independent cross section of DM elastic 
   scattering off with a nucleon and the DM pair annihilation cross section to a bottom quark pair at the present Universe. 
These cross sections are constrained by the results from the direct and indirect DM detection experiments, respectively. 
In the RS brane-world cosmology, the allowed parameter region is more severely constrained 
   than the standard Big-Bang cosmology case. 
   We find no allowed parameter space when we set the transition temperature to be its model-independent lower bound of $T_t \simeq 1$ MeV. 
On the other hand, the GB brane-world cosmological effect enlarges the allowed model parameter region, 
  and almost entire region for $m_D \gtrsim m_h/2$ is found to be consistent with all current experimental results.  
  %(as long as $\lambda$ remains in the perturbative regime). 
We find the direct DM detection constraints to be the most severe, which may indicate that the future direct detection experiments has high potential to discover WIMP DM in the near future. 
Once discovered, we can determine the transition temperature in the GB brane-world cosmology.

%It is worth investigating the GB brane-world cosmological effects on other physics 
%  such as inflation and baryogenesis. 
%We leave it for future work.   

%%%%%%%%%%%%%%%%%%%%%%%%%%%%%%%%%%
\section*{Acknowledgments}
%%%%%%%%%%%%%%%%%%%%%%%%%%%%%%%%%%
This work is supported in part by 
  the United States Department of Energy Grant DE-SC-0012447 (T.L. \& N.O.). 
  
\bibliographystyle{utphysII}
\bibliography{References}

\end{document}